\documentclass[superscriptaddress,aps,prb,preprint]{revtex4-2}
\usepackage{bm}
\usepackage{dcolumn}
\usepackage{graphicx}
\usepackage{hyperref}
\usepackage{mathrsfs}
\usepackage{multirow}
\usepackage{rotating}
\usepackage{url}
\usepackage{color}
\newcommand{\etal}{{\em et al}.\ }

\newcommand{\note}[1]{{{{#1}}}} 

\begin{document}


\title{Shift current response in elemental two-dimensional ferroelectrics}

\author{Zhuang Qian}
\affiliation{Zhejiang University, Hangzhou, Zhejiang 310058, China}
\affiliation{Key Laboratory for Quantum Materials of Zhejiang Province, Department of Physics, School of Science and
Research Center for Industries of the Future, Hangzhou Zhejiang 310030, China}
\author{Jian Zhou}
\affiliation{Center for Alloy Innovation and Design, State Key Laboratory for Mechanical Behavior of Materials, Xi’an Jiaotong University, Xi’an 710049, China}
\author{Hua Wang}
\affiliation{Zhejiang University, Hangzhou, Zhejiang 310058, China}
\author{Shi Liu}
\email{liushi@westlake.edu.cn}
\affiliation{Key Laboratory for Quantum Materials of Zhejiang Province, Department of Physics, School of Science and
Research Center for Industries of the Future, Hangzhou Zhejiang 310030, China}
\affiliation{Institute of Natural Sciences, Westlake Institute for Advanced Study,Hangzhou, Zhejiang 310024, China}

%

\date{\today}


\begin{abstract}
A bulk material without inversion symmetry can generate a direct current under illumination. This interface-free current generation mechanism, referred to as the bulk photovoltaic effect (BPVE), does not rely on $p$-$n$ junctions.  Here, we explore the shift current generation, a major mechanism responsible for the BPVE, in single-element two-dimensional (2D) ferroelectrics represented by phosphorene-like monolayers of As, Sb, and Bi. The strong covalency, small band gap, and large joint density of states afforded by these elemental 2D materials give rise to large shift currents, outperforming many state-of-the-art materials. We find that the shift current, due to its topological nature, depends sensitively on the details of the Bloch wave functions. It is crucial to consider the electronic exchange-correlation potential beyond the generalized gradient approximation as well as the spin-orbit interaction in density functional theory calculations to obtain reliable frequency-dependent shift current responses. 
\end{abstract}

\pacs{
}
\maketitle

\newpage

\section{Introduction}
High-performing photoelectric conversion is essential to the solar cell technology. Traditional photovoltaic cells based on $p$-$n$ junctions need the built-in electric field at the interface to separate electron-hole pairs and the efficiency is constrained by the Shockley–Queisser limit~\cite{Shockley61p510}. The bulk photovoltaic effect (BPVE) is the direct conversion of solar energy into direct current (DC), which has been considered as a promising alternative source of photocurrent~\cite{Sturman21book, Fridkin01p654, Grinberg13p509, Butler15p838, Tan16p16026, Spanier16p611}. As the name suggests, the presence of the BPVE does not require a complicated interface that often demands precisely controlled heterostructure fabrication process to minimize unwanted impurities and electric resistance. Single-phase bulk materials with  broken inversion symmetry can generate steady photocurrent and above-band-gap photovoltage under uniform illumination in the absence of external bias~\cite{Zenkevich14p161409, Nakamura17p281, Pal18p8005, Osterhoudt19p471, Huang2020p33950}, potentially enabling the implementation of the whole bulk material for photoelectric conversion~\cite{Zhang19p349, Yang10p143, Li21p5896, Burger19p5588}.

In the clean limit, the BPVE response can be obtained from the quadratic response theory using the density matrix method~\cite{Kraut79p1548, vonBaltz81p5590} or from the perspective of divergent susceptibilities~\cite{Aversa95p14636, Sipe00p5337, Fregoso19p064301} where the light is treated as a classical electromagnetic field interacting with Bloch  electrons. Ferroelectrics intrinsically exhibit the BPVE because of the fundamental requirement of spontaneous inversion symmetry breaking. Shift current is one of the most important mechanisms responsible for the BPVE~\cite{Sturman20p407, Panday19p195305, Dai21p177403}, and was first observed in ferroelectrics experimentally~\cite{Glass74p233}. As a zero-bias topological photocurrent~\cite{Barik20p045201, Ahn20p041041}, shift current is intimately related to the change in the phases of Bloch wave functions during the photoexcitation of an electron from the valence to the conduction band~\cite{Nastos06p035201}. A large shift current is desirable for photoelectric conversion.  Previous experimental and theoretical investigations of shift current in a wide range of noncentrosymmetric materials systems such as bulk perovskite ferroelectrics~\cite{Young12p116601, Young12p236601, Young15p054004, Tan19p085102}, two-dimensional (2D) materials~\cite{Range17p067402, Wang19p9743, Tiwari22p435404, Zhang19p3783, Xu21p4330, Xiao22p138}, nanotubes~\cite{Kim2022p3237}, conjugated polymers~\cite{Liu17p6500}, and topological insulators~\cite{Tan16p237402, Xu21p31, Xu22p111, Xu21p2101508, Ji19p955} have led to two general design principles for enhancing the current density.
First, low-dimensional materials with large joint density of states (JDOS) tend to present large shift current responses upon  photoexcitation~\cite{Cook17p14176}. The delocalization of electronic states is another important parameter that affects the shift current magnitude, and covalently bonded materials characterized by large long-range electron hopping amplitudes could give rise to large shift currents~\cite{Young12p116601, Tan16p16026, Liu17p6500}. 

Recently, a family of phosphorene-like 2D elemental group-V (As, Sb, and Bi) monolayers was predicted to possess spontaneous switchable in-plane polarization~\cite{Xiao18p1707383} arising from the out-of-plane atomic-layer buckling. We propose that single-element 2D ferroelectrics are promising candidates to support large shift currents because of the strong covalency intrinsic to the homoelement bonding and pronounced singularities in the density of states common to 2D materials.  In this work, we explore the shift current in these newly 
predicted elemental ferroelectrics with density functional theory (DFT) calculations and find that they can generate large shift currents over a wide range of wavelengths including the visible light spectrum. In addition, our work highlights the importance of spin-orbit coupling (SOC) and electronic correlation effect on the shift current spectrum even in 2D systems containing light elements.

\section{Results and Discussion}
\subsection{Structure}
\begin{figure}[h]
    \centering
    \includegraphics[width=0.6\textwidth]{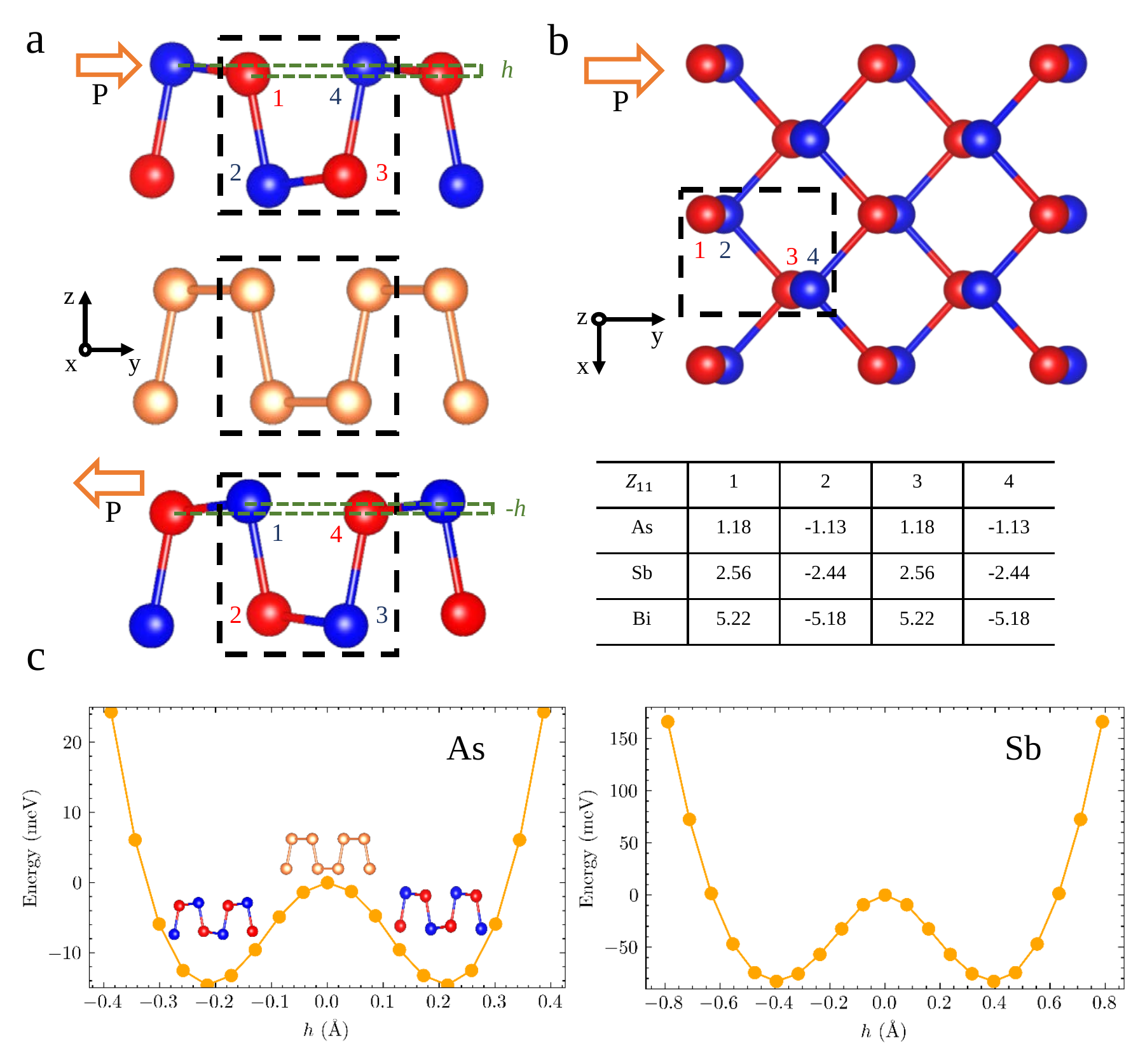}
    \caption{Ferroelectric single-element group-V monolayers. Schematics of (a) side view and (b) top view of crystal structures of single-element group-V monolayer with in-plane polarization along the $y$ axis. The inversion symmetry breaking results from the spontaneous atomic layer buckling denoted as $h$.
 The atoms are colored based on the sign of the Born effective charge ($Z_{11}$) reported in the table (blue for negative and red for positive). (c) Energy evolution as a function of buckling height $h$ in monolayer As monolayer (left) and monolayer Sb (right), revealing a double well potential.}
    \label{structure}
\end{figure}

The structure of single-element group-V monolayer is displayed in Fig.~\ref{structure}. The puckered lattice structure without centrosymmetry (space group $Pmn2_1$) can be viewed as a distorted phosphorene-like structure. The out-of-plane atomic buckling causes charge accumulations at the outmost group-V atoms, leading to spontaneous in-plane polarization along the $y$ axis~\cite{Xiao18p1707383}. We compute the Born effective charges (see inset table in Fig.~\ref{structure}) and confirm the buckling-induced charge accumulation/depletion mechanism. The dynamic stability of ferroelectric group-V monolayer has been confirmed by the computed phonon spectrum that has no imaginary phonon modes over the whole Brillouin zone~\cite{Xiao18p1707383}. Xu~\etal~recently simulated the polarization-electric field  hysteresis loop for monolayer As~\cite{Xu22p195418}, and the predicted in-plane spontaneous polarization is 0.42$\times$10$^{-10}$ C/m, comparable with ferroelectric monolayer SnTe~\cite{Wan17p132904}. We compute the energy evolution as a function of buckling parameter $h$ (Fig.~\ref{structure}a) in monolayer As and Sb, respectively, each revealing a double well potential landscape, typical for a ferroelectric phase (Fig.~\ref{structure}c). It is worth noting that group-V monolayers, similar to phosphorene, have already been experimentally synthesized for Sb and Bi~\cite{Wang06p233105, Bianchi12p155431, Nagao04p105501}. 
We only consider in-plane shift currents ($\sigma^{xbc}$ and $\sigma^{ybc}$) in this work. Since the mirror symmetry $\mathcal{M}_x: x \to -x$ leaves the monolayer invariant,  only three components of the response tensor, $\sigma^{yxx}$, $\sigma^{yyy}$, and $\sigma^{yzz}$, can be nonzero due to the symmetry constraint that holds at any photon frequency. Here, we focus on the responses under incident light perpendicular to the 2D sheet, namely, $\sigma^{yxx}$ and $\sigma^{yyy}$.


\subsection{Monolayer arsenic}
\begin{figure}[h]
    \centering
    \includegraphics[width=0.8\textwidth]{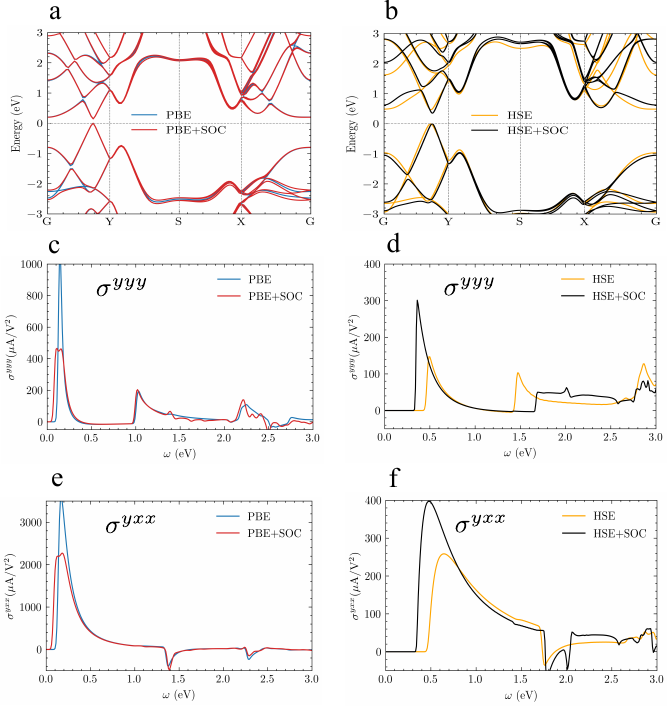}
    \caption{Electronic band structures and shift current spectra of monolayer As. Electronic band structures and shift current spectra of monolayer As. Band structures computed with (a) PBE and PBE+SOC and (b) HSE and HSE+SOC. $\sigma^{yyy}$ spectra computed with (c) PBE and PBE+SOC and (d) HSE and HSE+SOC. $\sigma^{yxx}$ spectra computed with (e) PBE and PBE+SOC and (f) HSE and HSE+SOC.}
    \label{SC_As}
\end{figure}

The DFT band structures and the corresponding shift current spectra for monolayer As computed with PBE, PBE+SOC, HSE, and HSE+SOC are presented in Fig.\ref{SC_As}, revealing several intriguing features. The band gap predicted by PBE is 0.15 eV, and the peak in the $\sigma^{yyy}$ spectrum exceeds 1000 $\mu$A/V$^2$ for photon energies near the band edge.
In comparison, HSE predicts a larger band gap of 0.47 eV, whereas the peak value of $\sigma^{yyy}$ drops considerably to only 150~$\mu$A/V$^2$. Similarly, the band-edge value of $\sigma^{yxx}$ estimated by PBE is more than 3000 $\mu$A/V$^2$ but the HSE predicts a much smaller value of 250 $\mu$A/V$^2$. Such pronounced reduction in the peak response is unexpected as the HSE band structure looks like a rigid shift of the PBE band structure without substantial changes in the band dispersions. This highlights the importance of treating the exchange-correlation potential beyond the semilocal approximation in DFT calculations. 

Moreover, the inclusion of the SOC effect, though inducing little impact on the band gap, causes drastic changes in the shift current spectra for both PBE and HSE. At the PBE level, the band gap reduction due to SOC is merely 0.01~eV, but the band-edge values of $\sigma^{yyy}$ and $\sigma^{yxx}$ computed with SOC reduce to 450 and 2200~$\mu$A/V$^2$ from their PBE values of 1000 and 3000$\mu$A/V$^2$, respectively; the PBE and PBE+SOC spectra are almost identical for higher photon frequencies. Interestingly, compared to the HSE values of 150 and 250~$\mu$A/V$^2$, the peak values of $\sigma^{yyy}$ and $\sigma^{yxx}$ estimated with HSE+SOC increase to 300 and 400~$\mu$A/V$^2$, respectively. In addition, the HSE and HSE+SOC spectrum profiles are notably different for both $\sigma^{yyy}$ and $\sigma^{yxx}$ in the frequency range between 1.5 and 3.0 eV.





\begin{figure}[h]
    \centering
    \includegraphics[width=0.8\textwidth]{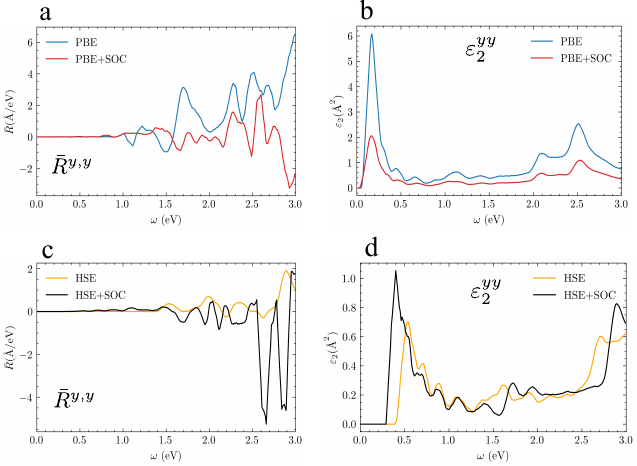}
    \caption{Shift vector and transition intensity in monolayer As. BZ-integrated shift vector ($\bar{R}^{y,y}$) and transition intensity ($\varepsilon_2^{yy}$) for $\sigma^{yyy}$ in monolayer As. $\bar{R}^{y,y}$ estimated with (a) PBE and PBE+SOC and (c) HSE and HSE+SOC. $\varepsilon_2^{yy}$ estimated with (b) PBE and PBE+SOC and (d) HSE and HSE+SOC. }
    \label{vector_As}
\end{figure}

To get a better understanding of the subtle changes in the shift current responses due to SOC, we analyze the spectrum of $\sigma^{yyy}$ by plotting two BZ-integrated quantities:  the energy averaged shift vector $\bar{R}^{a,b}$ and the transition intensity $\varepsilon_2^{bb}$ defined as
\begin{equation}
    \bar{R}^{a,b}(\omega) = \sum_{k} \sum_{nm} f_{nm}R_{nm}^{a,b}\delta(\omega_{mn}-\omega)
    \label{R}
\end{equation}
and
\begin{equation}
    \varepsilon_2^{bb}(\omega) = \sum_{k} \sum_{nm} r_{nm}^{b}r_{mn}^{b}\delta(\omega_{mn}-\omega),
    \label{epsilon}
\end{equation}
respectively. For a given photon frequency $\omega$, $\bar{R}^{a,b}(\omega)$ is a measure of aggregate contributions of shift vectors $R_{nm}^{a,b}$ at all $k$ points in the BZ; $\varepsilon_2^{bb}(\omega)$ reflects the photon absorption strength.
By comparing the spectra of $\bar{R}^{y,y}$ and $\varepsilon_2^{yy}$ computed with PBE, PBE+SOC, HSE, and HSE+SOC (Fig.~\ref{vector_As}), we find that the integrated shift vectors are comparable at the low-frequency region but these four methods predict rather different transition intensities near the band edge. Specifically, PBE predicts a larger magnitude of $\varepsilon_2^{yy}$ than PBE+SOC, whereas HSE yields a lower transition intensity than HSE+SOC. Therefore, the SOC effect suppresses the transition at the PBE level but promotes the transition in HSE. According to eq.~\ref{epsilon}, the magnitude of $\varepsilon_2^{yy}$ depends on the interband Berry connections between every conduction and valence band pair that has the energy difference matching the energy of the incident light. The presence of SOC, regardless the strength, will lift the spin degeneracy and lead to level anticrossing at some $k$ points. These changes in electronic bands albeit localized in the momentum space could strongly affect the interband Berry connections. Therefore, the pronounced SOC effect on the spectrum of $\varepsilon_2$ and then $\sigma$ is a manifestation of the topological nature of shift current that exhibits highly nontrivial dependence on the Berry connections of a bundle of Bloch bands (valence and conduction bands relevant to photon excitation). 

\begin{figure}[h]
    \centering
    \includegraphics[width=0.8\textwidth]{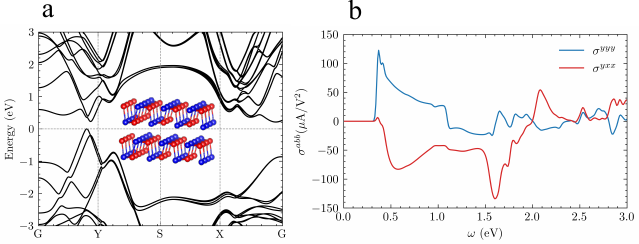}
    \caption{Band structure and shift current in bilayer As. (a) Electronic band structure and (b) shift current $\sigma^{yyy}$ and $\sigma^{yxx}$ spectra for bilayer As calculated with HSE+SOC. The inset in (a) shows the structure of bilayer As.
}
    \label{multi_bandsc}
\end{figure}

Previous studies have demonstrated that the shift current response of 2D materials can exist a significant dependence on the layer number~\cite{Strasser22p4145, Mu23p013001, Li21p5896}. We examine the layer stacking impact on the response functions. Our HSE+SOC calculations indicate that bilayer As remains polar and possesses a small band gap of 0.36 eV (Fig.~\ref{multi_bandsc}a), whereas trilayer As becomes centrosymmetric and metallic. The shift current spectrum is presented in Fig.~\ref{multi_bandsc}b, revealing a peak value of 125 $\mu$A/V$^2$ for $\sigma^{yyy}$ and $-135$ $\mu$A/V$^2$ for $\sigma^{yxx}$; these magnitudes are smaller than those in monolayer, consistent with those observed experimentally in CuInP$_2$S$_6$ where the photocurrent density decreases drastically when the thickness exceeds $\approx$40 nm~\cite{Li21p5896}.

We believe HSE+SOC is the most reliable method among the ones employed in the current work and most previous works, and the predicted peak shift current response in monolayer As is 400~$\mu$A/V$^2$ for $\sigma^{yxx}$, much higher than previous reports for 2D materials, i.e., 100~$\mu$A/V$^2$ in GeS. This highlights the potential of elemental ferroelectric 2D materials for photoelectric conversion. 


\subsection{Monolayer antimony}
\begin{figure}[h]
    \centering
    \includegraphics[width=0.8\textwidth]{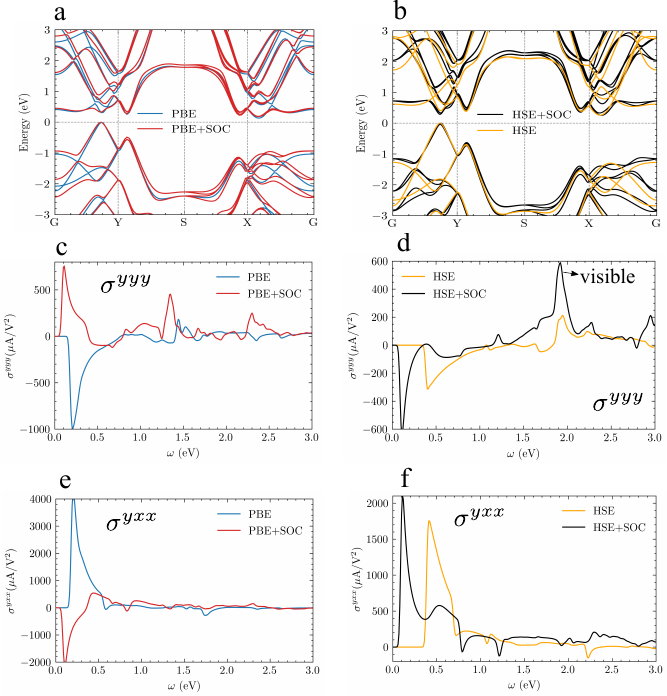}
    \caption{Electronic band structures and shift current spectra of monolayer Sb. Electronic band structures and shift current spectra of monolayer Sb. Band structures computed with (a) PBE and PBE+SOC and (b) HSE and HSE+SOC. $\sigma^{yyy}$ spectra computed with (c) PBE and PBE+SOC and (d) HSE and HSE+SOC. $\sigma^{yxx}$ spectra computed with (c) PBE and PBE+SOC and (d) HSE and HSE+SOC. }
    \label{SC_Sb}
\end{figure}

The band structures and shift current spectra of monolayer Sb computed with four different methods are displayed in Fig.~\ref{SC_Sb}. The direct band gap values based on the band structures plotted along high-symmetry lines are 0.31, 0.32, 0.41, and 0.29~eV for PBE, PBE+SOC, HSE, HSE+SOC, respectively. Despite yielding comparable band structures, these four methods predict distinct shift current spectra. Similar to the case of monolayer As, we observe a reduction of band-edge response magnitude of $\sigma^{yyy}$ ($\sigma^{yxx}$) from $-1000$ (4000)~$\mu$A/V$^2$ in PBE (Fig.~\ref{SC_Sb}c, e) to $-300$ (1700)~$\mu$A/V$^2$ in HSE (Fig.~\ref{SC_Sb}d, f), reaffirming the importance of including exact exchange in DFT calculations. 

\begin{figure}[h]
    \centering
    \includegraphics[width=0.8\textwidth]{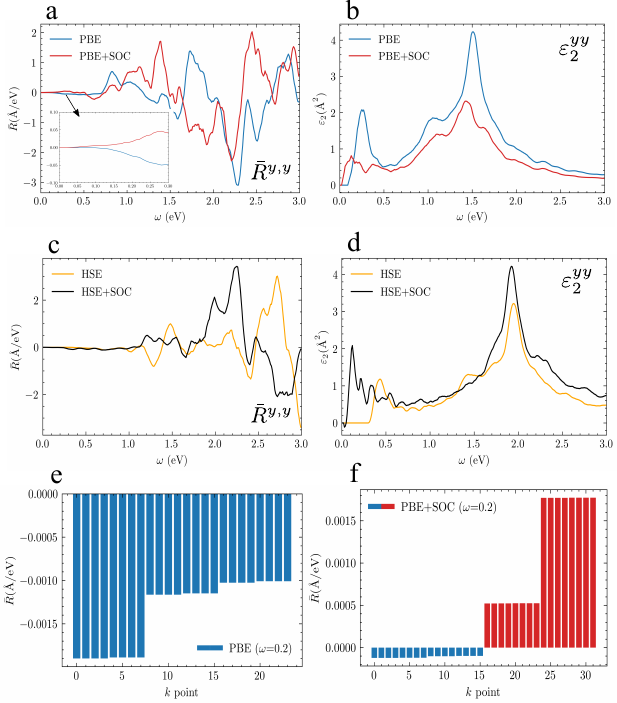}
    \caption{Shift vector and transition intensity in monolayer Sb. BZ-integrated shift vector ($\bar{R}^{y,y}$) and transition intensity ($\varepsilon_2^{yy}$) for $\sigma^{yyy}$ in monolayer Sb. $\bar{R}^{y,y}$ estimated with (a) PBE and PBE+SOC and (c) HSE and HSE+SOC. $\varepsilon_2^{yy}$ estimated with (b) PBE and PBE+SOC and (d) HSE and HSE+SOC. (e) and (f) are $k$-reolsved shift vector calculate at $\omega=0.2$ for PBE and PBE+SOC.}
    \label{vector_Sb}
\end{figure}

The inclusion of the SOC effect completely changes the spectrum profiles of $\sigma^{yyy}$ and $\sigma^{yxx}$ at the PBE level (Fig.~\ref{SC_Sb}c, e). The most striking result is the reversal of the current direction for low-frequency excitations. For example, PBE predicts a current running against the polarization ($\sigma^{yyy}<0$) but PBE+SOC predicts a current flowing along the polarization ($\sigma^{yyy}>0$). The sign of the shift current is determined by the integrated shift vector as defined in eq.~\ref{R}. Indeed, as shown in Fig.~\ref{vector_Sb}a, the sign of $\bar{R}^{y,y}$ is the same as the sign of $\sigma^{yyy}$, and SOC causes a sign change. For a given photon energy $\omega$, the value of $\bar{R}^{a,b}$ depends on the topological quantity $R_{nm}^{a,b}$ at every $k$ point in the BZ that supports resonant excitation. Following our previous argument regarding the effect of SOC on the transition intensity, the intraband Berry connection ($\mathcal{A}$) of some bands may be altered drastically by SOC, particularly around the $k$ point where SOC induces a hybridization gap. Therefore, it is physically plausible to have a sign changing situation due to SOC, as demonstrated in the case of monolayer Sb when treated with PBE. We further compute $k$-resolved $\bar{R}^{y,y}$ at a photon energy of 0.2~eV with PBE and PBE+SOC, respectively, with results ploted in Fig.~\ref{vector_Sb}e and f. It is found that all $k$-resolved $\bar{R}^{y,y}$ computed with PBE have negative values whereas those computed with PBE+SOC become mostly positive.

\begin{figure}[h]
    \centering
    \includegraphics[width=0.8\textwidth]{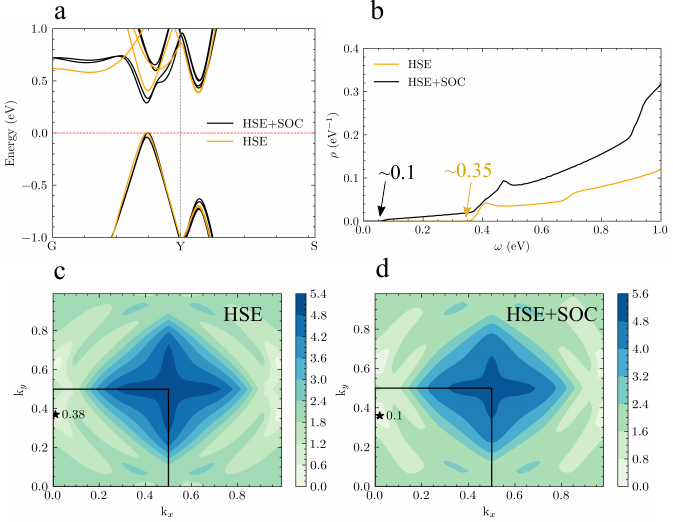}
    \caption{Analysis of the electronic structure of monolayer Sb. (a) Zoomed-in band structures calculated with HSE and HSE+SOC. (b) Joint density of states. Contour plots of the energy difference between the highest valence band and the lowest conduction band over the whole 2D BZ obtained with (c) HSE and (d) HSE+SOC. The star symbol labels the direct band gap.}
    \label{bandjdos_Sb}
\end{figure}

At the HSE level, SOC causes a redshift of the band-edge peak to $\approx0.1$~eV for both $\sigma^{yyy}$ and $\sigma^{yxx}$ (Fig.~\ref{SC_Sb}d, f), though the band gap value taken from the HSE+SOC band structures is 0.29~eV, higher than the onset photon energy. We further decompose $\sigma^{yyy}$ into $\bar{R}^{y,y}$ and $\varepsilon_2^{yy}$ and find that the magnitudes of $\bar{R}^{y,y}$ computed with HSE and HSE+SOC are comparable at the low-frequency region (Fig.~\ref{vector_Sb}c).  However, the $\varepsilon_2^{yy}$ spectrum of HSE+SOC has a sharp peak at a much lower frequency of 0.1~eV (Fig.~\ref{vector_Sb}d), consistent with the HSE+SOC spectrum of $\sigma^{yyy}$ (Fig.~\ref{SC_Sb}d). This seems to suggest forbidden optical excitations with in-gap photon frequencies. To resolve this puzzle, we perform a diagnostic analysis on the electronic structures of monolayer Sb obtained with HSE and HSE+SOC. The zoomed-in band structures along $\Gamma$-Y-S presented in Fig.~\ref{bandjdos_Sb}a show that the inclusion of SOC gives a smaller band gap of 0.29 eV than the  HSE band gap of 0.4~eV, and breaks the spin degeneracy. We compute the JDOS defined as 
\begin{equation}
    \rho(\omega) = \int\frac{d\bm{k}}{8\pi^3}\delta(\omega_{nm}-\omega)
\end{equation}
with results plotted in Fig.~\ref{bandjdos_Sb}b. The onset frequency of the HSE JDOS is 0.4~eV, as expected from the HSE band gap. However, the JDOS computed with HSE+SOC acquires nonzero values staring at a lower frequency of 0.1~eV. As the JDOS is obtained by integrating over the whole BZ while the band structure only shows the band energies along the high-symmetry BZ boundary paths, the JDOS of HSE+SOC hints at a smaller gap at a generic $k$-point. We thus map out the energy difference between the highest valence band and the lowest conduction band over the whole BZ (Fig.~\ref{bandjdos_Sb}c-d) with HSE and HSE+SOC, respectively. Indeed, HSE gives a gap of 0.38~eV at a $k$ point very close to the high-symmetry line $\Gamma$-Y (Fig.~\ref{bandjdos_Sb}c); HSE+SOC yields a band gap of 0.1~eV at a generic $k$-point ([0.04, 0.35] in reduced coordinates) away from the zone boundary (Fig.~\ref{bandjdos_Sb}d), consistent with the onset photon frequency for $\sigma^{yyy}$ and $\sigma^{yxx}$ predicted by HSE+SOC (Fig.~\ref{SC_Sb}d, f). It is noted that the peak value of $\sigma^{yxx}$ in monolayer Sb estimated with HSE+SOC reaches 2000~$\mu$A/V$^2$, even higher than monolayer As. Moreover, the value of $\sigma^{yyy}$ reaches 600~$\mu$A/V$^2$ at $\omega=1.8$~eV, suitable for visible light absorption. 

It is noted that we have compared the shift current spectrum and JDOS of monolayer GeS, a model material for investigating BPVE in 2D, with monolayer Sb. The peak shift current in GeS is $\approx$100 $\mu$A/V$^2$, one order of magnitude smaller than that in monolayer Sb. We find that monolayer Sb exhibits much larger JDOS than GeS. Nevertheless, because GeS has a much larger band gap than Sb and the shift current scales inversely with the band gap according to velocity gauge formalism~\cite{Kraut79p1548}, we suggest that the giant shift current in monolayer Sb could be attributed to multiple factors including small band gap and large JDOS.

\subsection{Monolayer bismuth}
\begin{figure}[h]
    \centering
    \includegraphics[width=0.8\textwidth]{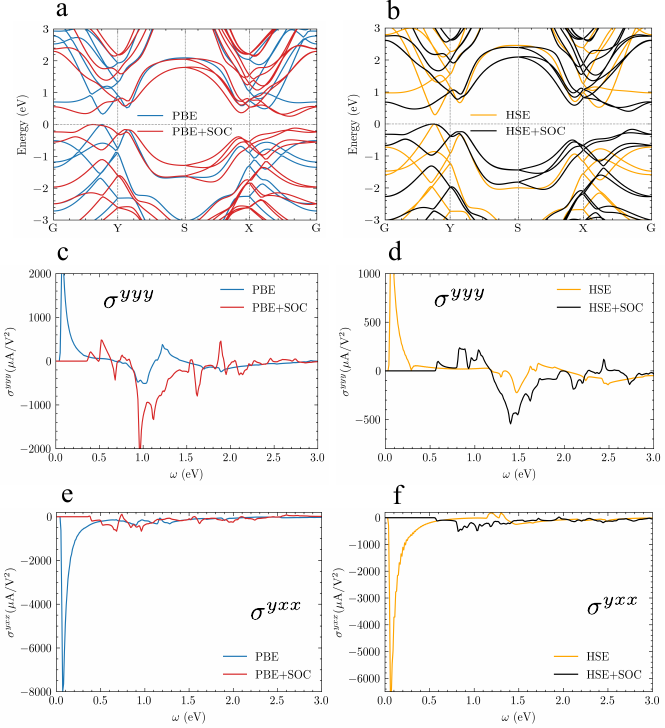}
    \caption{Electronic band structures and shift current spectra of monolayer Bi. Band structures computed with (a) PBE and PBE+SOC and (b) HSE and HSE+SOC. $\sigma^{yyy}$ spectra computed with (c) PBE and PBE+SOC and (d) HSE and HSE+SOC. $\sigma^{yxx}$ spectra computed with (c) PBE and PBE+SOC and (d) HSE and HSE+SOC.}
    \label{SC_Bi}
\end{figure}

\begin{figure}[h]
    \centering
    \includegraphics[width=0.8\textwidth]{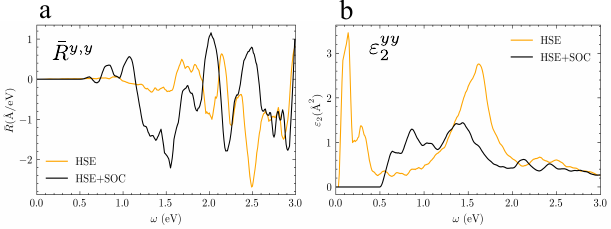}
    \caption{Shift vector and transition intensity in monolayer Bi. BZ-integrated (a) shift vector and (b) transition intensity for $\sigma^{yyy}$ in monolayer Bi computed with HSE and HSE+SOC.}
    \label{vector_Bi}
\end{figure}

Since Bi has a larger atomic number, the SOC effect is more pronounced in monolayer Bi as demonstrated from the notably different band structures between PBE (HSE) and PBE+SOC (HSE+SOC), as shown in Fig.~\ref{SC_Bi}a, b.
Without SOC, PBE and HSE predict similar spectrum profiles of $\sigma^{yyy}$ and $\sigma^{yxx}$ and large band-edge responses. The magnitude of the peak response of $\sigma^{yxx}$ computed with HSE is close to 6500~$\mu$A/V$^2$. In comparison, the spectra obtained with SOC are qualitatively different. In general, the spectra of PBE+SOC and HSE+SOC share similar peak structures with the latter predicting lower peak values (Fig.~\ref{SC_Bi}c-f). For example, PBE+SOC predicts a main peak of $-2000$~$\mu$A/V$^2$ at 1.0~eV for $\sigma^{yyy}$ while the HSE+SOC spectrum of $\sigma^{yyy}$ has the highest peak of $-500$~$\mu$A/V$^2$ located at 1.4~eV. We note that the band-edge response is no longer the strongest. The spectrum of $\varepsilon_2^{yy}$ (Fig.~\ref{vector_Bi}b) confirms that the SOC interaction reduces the band-edge photon absorption than that in HSE. Given the substantially different band structures predicted by HSE and HSE+SOC, it is not surprising to obtain drastically different spectra of $\bar{R}^{y,y}$ (Fig.~\ref{vector_Bi}a).
For monolayer Bi, the inclusion of SOC is crucial to acquire correct shift current spectrum.



\subsection{Strain effect}

\begin{figure}[h]
    \centering
    \includegraphics[width=0.8\textwidth]{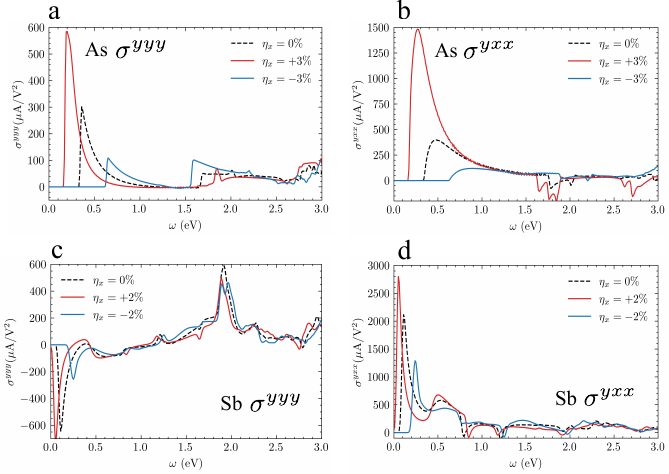}
    \caption{Shift vector and transition intensity in monolayer Bi. Uniaxial strain dependence of (a)(c) $\sigma^{yyy}$ and (b)(d) $\sigma^{yxx}$ of monolayer As and Sb.}
    \label{As_strain}
\end{figure}

Because reduced-dimensional structures can sustain much larger strains than their bulk counterparts, it has become common to use strain to modulate the structural, electronic, and optical properties of 2D materials~\cite{Xu20p3141, Li20p1151, Zhao20p046801, Wei18p1095}. Here, we explore the effect of uniaxial strain ($\eta_x$) on the shift current response. Taking monolayer As for example, we find that a tensile strain along the $x$ direction can effectively reduce the band gap and promote the current density. As shown in Fig.~\ref{As_strain}, a 3\% tensile strain enhances both $\sigma^{yyy}$ and $\sigma^{yxx}$ by nearly 4-fold. In contrast, a compressive strain ($\eta_x=-3$~\%) increases the band gap hence reduces the band-edge response. Interestingly, the peak of $\sigma^{yyy}$ at a higher photon energy of 1.6~eV gets enhanced by the compressive strain of $-3$\%. Similar tensile strain-promoted response is also found in monolayer Sb. In particular, the magnitude of $\sigma^{yxx}$ gets enhanced to 2800~$\mu$A/V$^2$ when stretching the monolayer by 2\% along $x$ (Fig.~\ref{As_strain}d).

\begin{figure}[h]
    \centering
    \includegraphics[width=0.8\textwidth]{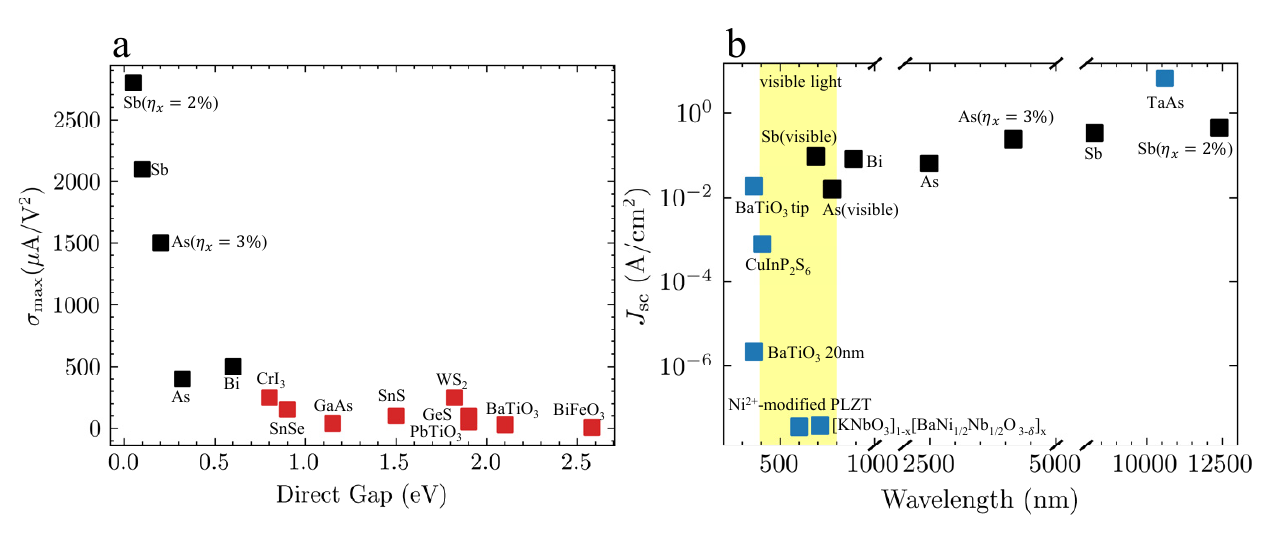}
    \caption{Comparison of shift current response and current density of different materials. (a) Peak value of shift current response~\cite{Young12p116601, Julen18p245143, Range17p067402, Wang17p115147, Zhang19p3783} and (b) current density~\cite{Li21p5896, Spanier16p611, Zenkevich14p161409, Grinberg13p509, Huang2020p33950, Pal18p8005, Osterhoudt19p471} of different materials assuming a light intensity of 0.1 W/cm$^2$. Besides strong band-edge responses to long-wavelength light, monolayer Sb and As can generate large shift currents in response to visible light illumination. }
    \label{context}
\end{figure}

Finally, we compare the shift current responses of different materials including bulk polar materials (e.g., PbTiO$_3$, BaTiO$_3$, and GaAs) and 2D materials (e.g., GeSe and CrI$_3$). As summarized in Fig.~\ref{context}a, the magnitude of the response tensor roughly scales inversely with the band gap, and the stretched monolayer Sb has the highest response of 2800~$\mu$A/V$^2$. Note that most previous results were based on PBE calculations, which generally overestimate the response. To compare with experimental data, we further estimate the shift photocurrent density based on the computed response tensor and a light intensity of 0.1 W/cm$^2$ following the method in ref.~\cite{Range17p067402}. It is evident from Fig.\ref{context}b that elemental 2D ferroelectrics represented by Sb, As, and Bi outperform conventional bulk ferroelectrics and 2D layered materials such as CuInP$_2$S$_6$ over a wide range of wavelengths including the visible light spectrum. It is worth noting that TaAs exhibits a strong response at a wavelength of $\approx$10000 nm, which is much larger than the wavelengths of visible light. 
Given TaAs is a Weyl semimetal, it is not surprising that its shift current response is large due to the gapless band dispersion and topological nature. The mid-infrared response is likely more relevant to applications such as optical detectors and sensors. In comparison, monolayers As and Sb exhibit large shift current response across a wide range of wavelengths including the visible light spectrum, which would be advantageous for utilizing most of the solar spectrum.
Additionally, since the peak responses of group-V elemental ferroelectrics are consistent with light-induced terahertz emission, the large current magnitude suggests their potential applications in terahertz source platforms.

In this work, we investigate the shift current responses in single-element two-dimensional ferroelectrics represented by monolayer As, Sb, and Bi in the space group of $Pmn2_1$ using first-principles density functional theory calculations. We find that PBE and HSE yield qualitatively different shift current spectra, demonstrating the importance of exact exchange potential for 
reliable predictions of optical responses. Moreover, the spin-orbit coupling, largely overlooked in previous studies, can substantially affect the magnitude, sign, and spectral profile of shift current even for light elements such as arsenic. This highlights the topological nature of shift current that has nontrivial dependence on both intraband and interband Berry connections of a bundle of valence and conduction bands. Regarding computational materials by design for new solar materials that can generate large shift currents, we suggest that it is essential to treat the electronic exchange-correlation interaction beyond the generalized gradient approximation and to include the spin-orbit interaction in density functional theory calculations. Based on the results predicted by HSE+SOC, we propose that elemental 2D ferroelectrics can support large shift currents, outperforming many state-of-the-art materials. This work unravels the potential of elemental 2D ferroelectrics for photovolatic and optoelectronic applications.

\section{Method}
We follow the formalism adopted in Ref.~\cite{Julen18p245143, Wang17p115147} to compute the photon frequency-dependent shift current spectrum. The shift current density ($J_2$) is regarded as a second-order optical response to the electromagnetic field $E$ of frequency $\omega$, 
\begin{equation}
    J_{2}^{a}=2\sum_{bc}\sigma^{abc}(0;\omega,-\omega)E^{b}(\omega)E^{c}(-\omega),
\end{equation}
where the third-rank tensor $\sigma^{abc}(0;\omega,-\omega)$ is the shift current response tensor; $a$, $b$, and $c$ are Cartesian indexes, and the index $a$ specifies the direction of the generated DC current, while $b$ and $c$ are polarization directions of an incident light. Within the length gauge, the shift current tensor is derived as
\begin{equation}\label{SCtensorabc}
    \sigma^{abc}(0;\omega,-\omega) = -\frac{i\pi e^3}{2\hbar^2}\int\frac{d\bm{k}}{8\pi^3}\sum_{nm}f_{nm}(r^{b}_{mn}r^{c}_{nm;a}+r^{c}_{mn}r^{b}_{nm;a})\delta(\omega_{mn}-\omega),
\end{equation}
where $n$ and $m$ are band indexes, and $\bm{k}$ is the wavevector of the Bloch wave function. $f_{nm}= f_n - f_m$ is the difference in the Fermi-Dirac occupation number between bands $n$ and $m$; $\omega_{nm} = \omega_n - \omega_m$ represents the band energy difference. $r^b_{nm} = i\langle n|\partial k^b |m \rangle$ is the dipole matrix element (interband Berry connection). $r^{b}_{nm;a}$ is the generalized derivative expressed as $r^{b}_{nm;a} = \frac{\partial r^b_{nm}}{\partial k^a}-ir^b_{nm}(\mathcal{A}^a_n - \mathcal{A}^a_m )$ with $\mathcal{A}^a_n = i\langle n|\partial k^a |n \rangle$ the intraband Berry connection.

Under a linearly polarized light ($b=c$), the shift current response tensor can be re-formulated into a more compact expression,
\begin{equation}\label{SCtensorlinearoptical}
    \sigma^{abb}(0;\omega,-\omega) = \frac{\pi e^3}{\hbar^2}\int\frac{d\bm{k}}{8\pi^3}\sum_{nm}f_{nm}R_{nm}^{a,b}r^b_{nm}r^b_{mn}\delta(\omega_{mn}-\omega),
\end{equation}
where $R_{nm}^{a,b} = \frac{\partial \phi^b_{nm}}{\partial k^a}+\mathcal{A}^a_n - \mathcal{A}^a_m $ is the shift vector with $\phi_{nm}$ being the phase factor of the dipole matrix element $r_{nm}^b = |r_{nm}^b|e^{-i\phi_{nm}^b}$. The shift vector has a unit of length and represents the average displacement of the coherent photoexcited carriers during their lifetimes. The $r^b_{nm}r^b_{mn}$ term measures the transition intensity which describes the optical absorption strength for the transition from band $m$ to band $n$. Therefore, the response tensor can be viewed as the product of shift vector and optical transition intensity.


The structural parameters of monolayer As, Sb, and Bi are optimized using \texttt{Quantum Espresso}~\cite{Giannozzi09p395502, Giannozzi17p465901} with Garrity-Bennett-Rabe-Vanderbilt (GBRV) ultrasoft pseudopotentials~\cite{Garrity14p446}. The exchange-correlation functional is treated within the generalized gradient approximation of Perdew-Burke-Ernzerhof (PBE) type~\cite{Perdew96p3865}. We use a plane-wave kinetic energy cutoff of 50 Ry, a charge density cutoff of 250 Ry, a 12$\times$12$\times$1 Monkhorst-Pack $k$-point mesh for Brillouin zone (BZ) integration, an ionic energy convergence threshold of 10$^{-5}$ Ry, and a force convergence threshold of 10$^{-4}$ Ry in structural optimizations.  Maximally localized Wannier functions to fit DFT electronic structures are obtained using the \texttt{Wannier90} code~\cite{Pizzi20p165902}, and then the shift current response tensor is calculated in the Wannier basis as described in ref.~\cite{Julen18p245143}. We use a numerical smearing parameter of 20 meV and a dense $k$-point grid of 1000$\times$1000$\times$1 to ensure the spectrum convergence. In addition, the 3D-like response is estimated by rescaling the calculated 2D response with the effective thickness of monolayer~\cite{Range17p067402}. The spin-orbit coupling (SOC) is taken into account at the fully relativistic level with norm-conserving pseudopotentials provided by the PseudoDoJo project~\cite{Van18p39}. We also compute the shift current using the Heyd-Scuseria-Ernzerhof (HSE) hybrid functional~\cite{Krukau06p224106} with a $8\times8\times1$ $q$-point grid during the self-consistent-field cycles. The HSE band structure is obtained via Wannier interpolation~\cite{Marzari12p1419} using \texttt{Wannier90} interfaced with \texttt{Quantum Espresso}. We employ the \texttt{Wannier Berri} code~\cite{Tsirkin21p33, Destraz20p5} to compute the shift vector and transition intensity (see discussions below) to analyze the shift current spectrum. \note{It is noted that all shift current calculations are based on the independent-particle approximation and do not take the exciton effect into account.}


\begin{acknowledgments}
Z.Q. and S.L. acknowledge the supports from Westlake Education Foundation. We acknowledge Dr. Jae-Mo Lihm for useful suggestions regarding the usage of Wannier90 and Wannier Berri. Z.Q. acknowledges the the help from Yudi Yang during the preparation of the manuscript. The computational resource is provided by Westlake HPC Center. J.Z. acknowledges National Natural Science Foundation of China under Grant No. 11974270.
\end{acknowledgments}

{\bf{Author Contributions}} S.L. conceived and led the project. Z.Q. performed calculations and data analysis. All authors contributed to the discussion and the manuscript preparation.

{\bf{Competing Interests}} The authors declare no competing financial or non-financial interests.

{\bf{Data Availability}} The data that support the findings of this study are included in this article and are available from the corresponding author upon reasonable request.

\bibliography{note}

\newpage

\newpage

\newpage

\newpage

\newpage

\newpage

\newpage

\newpage

\newpage

\newpage

\newpage

\end{document}